\documentclass[epj, nopacs]{svjour}
\usepackage{graphics}
\usepackage{graphicx}
\usepackage{amsmath,bm}
\usepackage{amssymb}
\usepackage{color}
\usepackage{epstopdf}
\newcommand{\ket}[1]{|#1\rangle}

\begin{document}
	\title{Deterministic quantum state transfer of atoms in a random magnetic field}
	\author{Bianca J. Sawyer\textsuperscript{1,2,}\thanks{bianca.j.sawyer@postgrad.otago.ac.nz} \and
		Matthew Chilcott\textsuperscript{1,2} \and
		Ryan Thomas\textsuperscript{1,2} \and
		Amita B. Deb\textsuperscript{1,2}  \and
		Niels Kj{\ae}rgaard\textsuperscript{1,2,}\thanks{nk@otago.ac.nz}}
	\institute{Department of Physics and QSO -- Centre for Quantum Science, University of Otago, Dunedin, New Zealand. \and The Dodd Walls Centre for Photonic and Quantum Technologies, New Zealand.}

	\date{Received: date / Revised version: date}
	
\abstract{We propose a method for transferring atoms to a target quantum state for a multilevel quantum system with sequentially increasing, but otherwise unknown, energy splitting. This is achieved with a feedback algorithm that processes off-resonant optical measurements of state populations during adiabatic rapid passage in real-time. Specifically, we reliably perform the transfer $\ket{F=2,m_F=2} \rightarrow \ket{1,1} \rightarrow \ket{2,1}$ for a sample of ultracold $^{87}$Rb in the presence of a random external magnetic field.}

\maketitle
\section{Introduction}
\label{intro}

Many quantum technologies rely on the ability to robustly prepare atoms and atom-like systems in given internal quantum states. One way to transfer population between two states separated by an energy $E$ is to apply resonant radiation of frequency $f=E/h$. This induces resonant Rabi flopping between the two levels and the final state is determined by the duration and intensity of the applied radiation. Obviously, fluctuations in the driving field directly translate into imperfect state preparation. Moreover, if the energy separation of the two levels depends on residual electromagnetic fields (as in the Zeeman and Stark effects), drift and fluctuations in these background fields lead to off-resonant driving fields, rendering the transfer imperfect.
	
Various types of errors encountered in connection to state preparation using a single radiation pulse can be mitigated through the use of composite pulses. Originally invented for robust spin manipulation in nuclear magnetic resonance (NMR) experiments \cite{Abragam1961}, the manipulation of a two-level system (equivalent to a spin $1/2$ particle) can be achieved through a sequence of pulses where the introduced errors tend to negate. A composite pulse is a special (discrete) case of irradiation being executed with arbitrary time dependence of both amplitude and phase (or equivalently, frequency) of the driving field.
	
\begin{figure}[!htbp]
	\centering
	\resizebox{1\columnwidth}{!}{\includegraphics[trim=0cm 0cm 0cm 0cm]{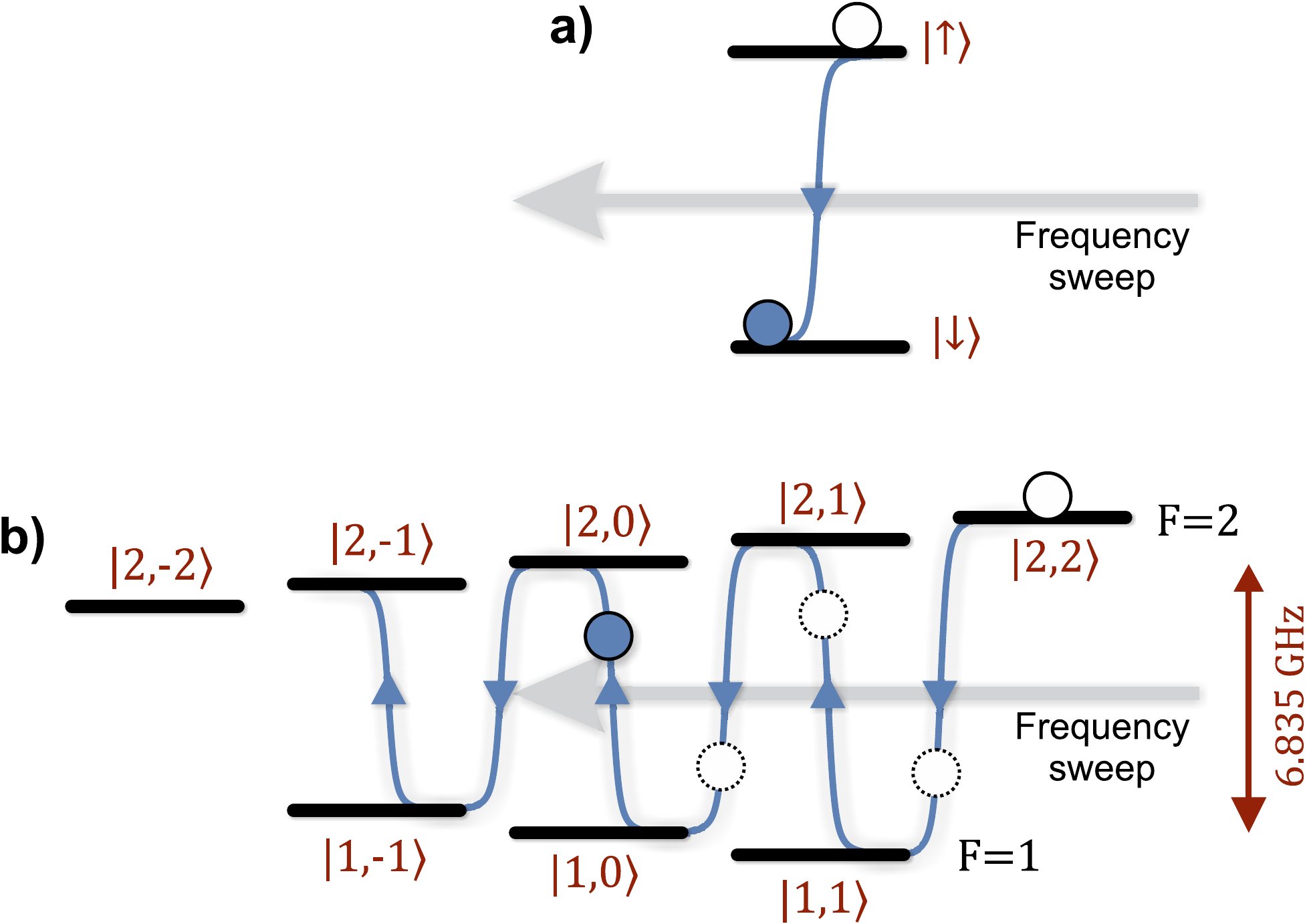}}
	\caption{a) Illustration of an ARP frequency sweep in a two-level system giving rise to population transfer from state $\ket{\uparrow}$ to state $\ket{\downarrow}$. b) The same ARP frequency sweep in a multi-level system, where the stopping frequency needs to be well-known in order to prepare atoms in a particular state. } \label{fig1:setstage}
\end{figure}

A widely used method for robust population transfer from one state to another is to use a frequency swept pulse, achieving so-called adiabatic rapid passage (ARP) \cite{Allen1987,Camparo1984}. In ARP the frequency of radiation sweeps adiabatically through the resonance from either above or below (the sign of the initial detuning is irrelevant), and, if this is performed sufficiently slowly, the initial populations are inverted  -- see Fig.~\ref{fig1:setstage}a. ARP has been proven to be very useful in preparing quantum states in the presence of small magnetic field fluctuations, as it is relatively insensitive to jitter in the transition frequency \cite{Kotru2014}. ARP thus relaxes demands on the accuracy of the frequency of the radiation field or, equivalently, accurate knowledge of the transition frequency of the two-level system.

The scenario for state preparation becomes more complex for a multilevel atomic system. In principle, for a system where the resonant frequency between levels successively increases, perfect state transfer between any pair of states can be achieved via ARP. As an example of this type of system, Fig.~\ref{fig1:setstage}b illustrates a microwave frequency ARP sweep across Zeeman sub-states within the $5^2S_{1/2}$ ground state of $^{87}$Rb. In the presence of an unknown magnetic field, however, the end point of the sweep required to prepare a particular state cannot be predicted as the Zeeman effect will modify the transition frequencies. If the applied frequency sweep undershoots or overshoots the desired transition, the final quantum state will be incorrect \cite{Malinovsky2001}. Obviously, knowledge about the evolution of the quantum state during the microwave sweep could be used to determine when the desired target state has been reached, and trigger termination of the external driving radiation field. Here we propose to use a measurement of the quantum state-dependent refractive index of an atomic ensemble to monitor the progress of transitions in real-time, and use the information gained to prepare a target quantum state with high fidelity.

The refractive index of an atomic sample can be measured via its interaction with laser light \cite{Bjorklund1980}. Dispersive probing is implemented with an off-resonant probe beam, which incurs a phase shift proportional to the refractive index of the sample but does not significantly alter the energy or quantum state of the atoms in the ensemble during the measurement process. Such off-resonant probing methods are thus well-suited to real-time monitoring, and are a relatively simple addition to a cold atoms experimental setup \cite{Lye1999}.

Off-resonant probing has been demonstrated previously in a variety of scenarios, for example to monitor forced evaporative cooling \cite{Sawyer2012} and spatial centre-of-mass oscillations \cite{Kohnen2011} of trapped atomic samples, the phase transition to Bose-Einstein condensate (BEC) \cite{Bason2018}, the dynamics of nonclassical collective spin states \cite{Smith2004} and Larmor precession \cite{Isayama1999}, and to perform gradient magnetometry \cite{Deb2013}. These dispersive monitoring techniques open up a pathway to measurement-based feedback control. Previously, feedback routines using off-resonant probing methods have been used in ultracold gas experiments to stabilize atom numbers \cite{Gajdacz2016} and to control the orientation of coherent spin states in real-time \cite{Vanderbruggen2013,Vanderbruggen2014,Gillett2010}.  Theoretical proposals have also considered quantum coherent feedback systems, where the controller is another quantum system \cite{Dong2010,Serafini2012}, for feedback cooling \cite{Steck2004,Mabuchi2008} and stabilisation \cite{Wilson2007,Szigeti2010} of a BEC, control of spinor BECs \cite{Wang2016} and machine learning \cite{Hentschel2010}.

As a proof-of-concept application of a closed-loop feedback system based on heterodyne dispersive probing measurements we demonstrate transfer of a sample of $^{87}$Rb to a particular Zeeman sub-state $\ket{F,m_F}$ using ARP. Our dispersive probe is sensitive to atoms in the $F=2$ state and insensitive to atoms in the $F=1$ state; thus, monitoring the dispersive signal while transitioning between these two manifolds gives us a distinct staircase of transitions with sequentially varying energy, see Fig.~\ref{fig1:setstage}b. By counting the rising and falling edges in the dispersive data we can then follow the progression through the quantum states and trigger termination of the ARP sweep when the desired state has been reached.

\section{Experimental Setup}
\label{sec:1}

\begin{figure*}
\centering
\includegraphics[width=\textwidth]{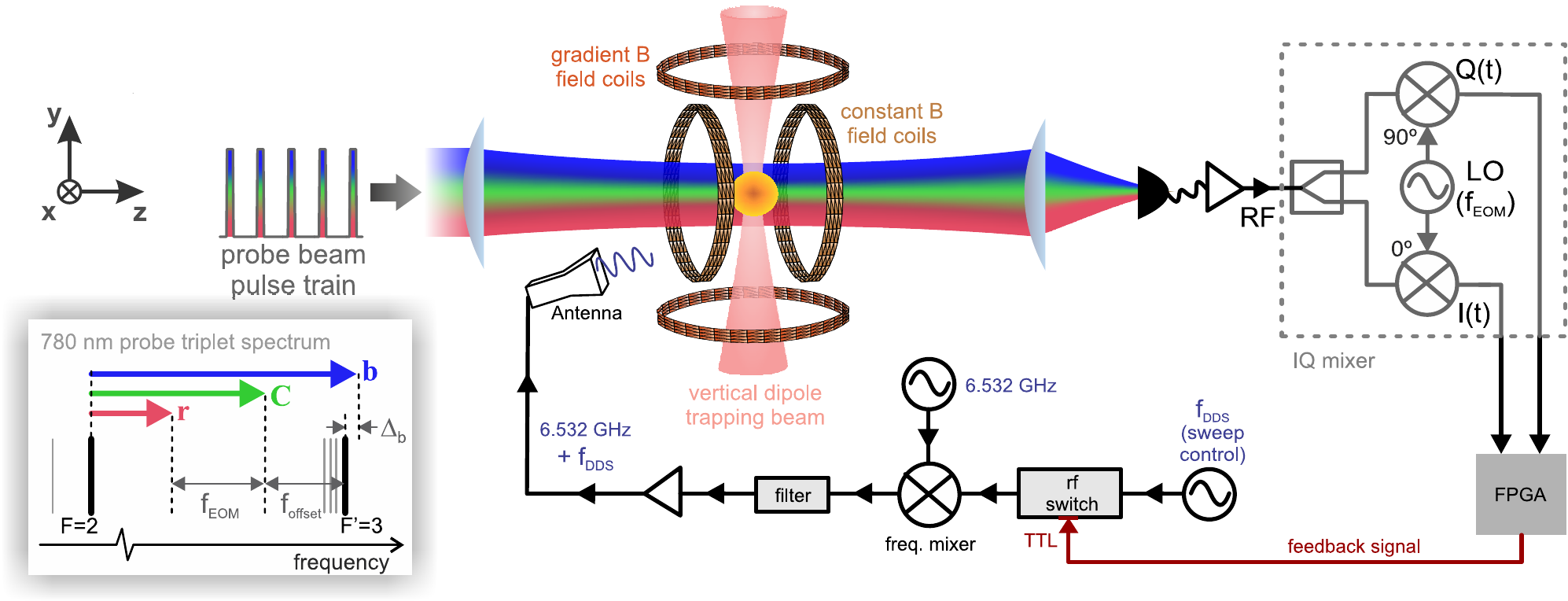}
\caption{Schematic of the experimental setup, showing a sample of $^{87}$Rb atoms trapped in a cross-beam dipole trap and being driven with near-resonant microwave frequency radiation from an antenna. The sample is illuminated along the $+z$-direction by a dispersive probing beam, which is then collected on a fast photodetector. The signal is amplified and demodulated to DC using an IQ mixer (on the right of the figure). The outputs are sampled by the FPGA board and processed by the feedback algorithm. The feedback loop is completed by a TTL output from the FPGA, which controls the rf switch, effectively turning off the transmission of near-resonant microwave radiation when the stop condition is met. Inset in bottom left corner: the trichromatic frequency spectrum of the probe, consisting of a carrier ($C$), and red ($r$) and blue ($b$) sidebands, relative to the $F=2 \rightarrow F'=3$ transition of the D2 line. The other (off-resonant) $F$ states ($F=1$ and $F'=0,1,2$) are indicated as narrower grey lines. Note that this figure is not to scale.}
\label{fig2:setup}
\end{figure*}

Figure~\ref{fig2:setup} shows a schematic of our experimental setup. We produce samples of ultracold $^{87}$Rb, with a temperature of $T\simeq1.5~\mu$K and atom number $N\simeq3\times10^6$, using a standard cold atom experiment \cite{Sawyer2012}. The atoms are initially prepared in the $5^2S_{1/2} \ket{F=2,m_F=2}$ hyperfine state in a crossed-beam optical dipole trap (1064~nm) \cite{Roberts2014}.

The splitting between the $F=1$ and $F=2$ hyperfine levels is $\sim6.8$~GHz. We address $F=1 \leftrightarrow F=2$ transitions with a driving field produced by mixing the 6.532~GHz single-frequency output of a waveform generator with a controllable $f_{\mathrm{DDS}}\sim300$~MHz signal. The latter is generated by a direct digital synthesiser (DDS), which has the desired frequency sweep profile uploaded to it in advance. The mixer output is filtered (with a 300~MHz wide passband, centred on 6.8~GHz) and the resulting $\simeq6.8$~GHz signal amplified before being transmitted to the atomic sample by a rectangular waveguide antenna. An external rf switch is used to switch the $f_{\mathrm{DDS}}$ input into the mixer on and off, which controls whether the driving field is incident on the sample.

During microwave frequency sweeps across the ground state Zeeman transitions, we use an off-resonant dispersive probe system \cite{Sawyer2017} to monitor the population of the $F=2$ hyperfine level. The dispersive probe is generated from 780~nm laser light detuned to $f_{\mathrm{offset}} = -3.30$~GHz below the $F = 2 \rightarrow F' = 3$ transition of the D2 line. This `carrier' component is passed through a fibre electro-optic phase modulator (EOM) driven at $\pm f_{\mathrm{EOM}}= \pm3.700$~GHz. This produces sidebands that co-propagate with the carrier to make up an interferometric, three-component optical frequency spectrum (the EOM is driven at a power such that all but the first-order sidebands are negligible). The inset on the left of Fig.~\ref{fig2:setup} shows schematically the probe frequency triplet relative to the $F=2 \rightarrow F'=3$ transition of the D2 line. The carrier ($C$) has an optical power of $\sim13~\mu$W and each of the two first-order sidebands contain $\sim1~\mu$W. The downshifted sideband (r) is far-red detuned $\Delta_{\mathrm{r}} = f_{\mathrm{offset}} - f_{\mathrm{EOM}} =-7.00$~GHz from the $2\rightarrow3$ transition, while the up-shifted (b) sideband has a comparatively small blue detuning of $\mathrm{\Delta}{\mathrm{b}} = f_{\mathrm{offset}} + f_{\mathrm{EOM}} = +400$~MHz.

We create a pulse train of probe laser light using an acousto-optic modulator (AOM). The pulses have a duration of 500~ns and are separated by 2.5~ms. The probe light is linearly polarized along the $x$ axis and propagated along the $z$ axis, being focused to a $28~\mu$m waist centred on the sample. As the probe triplet passes though the atomic cloud the $b$ component acquires a phase shift dependent upon the $F = 2$ population of the cloud. The beam is then focused onto a $4.2$~GHz bandwidth fibre-coupled ac photodetector, where the three frequency components combine to produce a beat signal at frequency $f_{\mathrm{EOM}}$. This signal is amplified and demodulated to DC using an IQ mixer (illustrated on the right of Fig.~\ref{fig2:setup}).

\begin{figure} \centering  
	\resizebox{1\columnwidth}{!}{%
		\includegraphics[width=\columnwidth, trim=0cm 0cm 0cm 0cm]{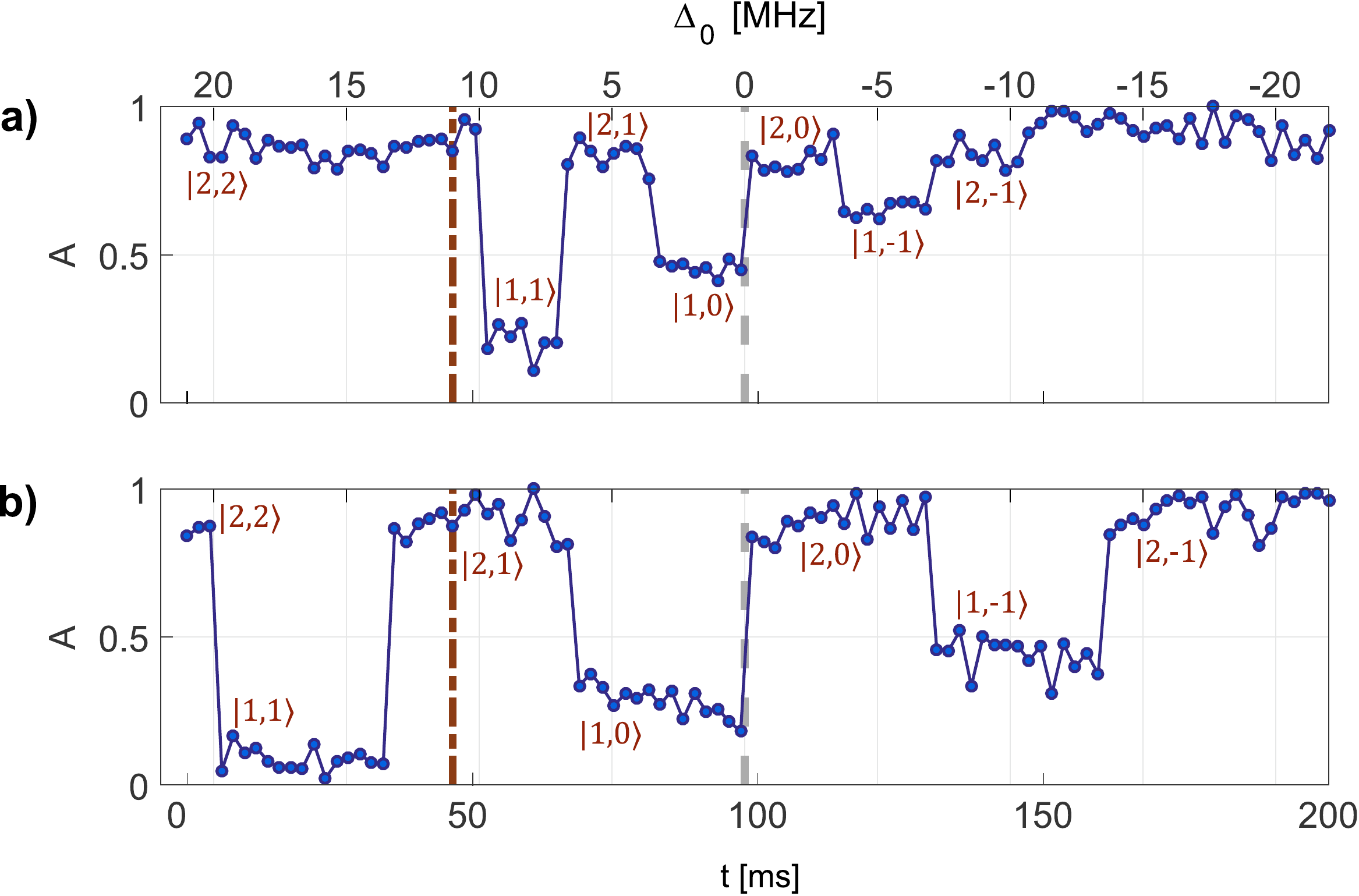}}
	\caption{a) Dispersive probe signal versus time as the frequency of the microwave driving field is swept linearly from $\mathrm{\Delta_{0}}=+21$~MHz to $\mathrm{\Delta_{0}}=-22$~MHz, in the presence of an external magnetic field, $B_z = 4.7$~G.
		The sample is initially prepared in the $\ket{2,2}$ state, and this sweep covers all six transitions (as illustrated schematically in Fig.~\ref{fig1:setstage}b). b) A repeat of the situation shown in a), but with external magnetic field $B_z = 9.3$~G.
		The dashed grey line shows the position of the zero-field $F=1 \leftrightarrow F=2$ transition, and the red dash-dotted line indicates where $\mathrm{\Delta_{0}}=11$~MHz.} \label{fig3:sweeps}
\end{figure}

Figure~\ref{fig3:sweeps}a shows the dispersive probing signal produced for an atomic sample addressed by microwave radiation sweeping across a 43~MHz frequency range centred on $f_{\mathrm{0}} = 6.83468261$~GHz \cite{Bize1999} over 200~ms, in the presence of an external magnetic field, $B_z = 4.7$~G. This frequency range is sufficiently wide to cover all possible $F=1 \leftrightarrow F=2$ transitions in the presence of an external magnetic field of up to $\pm 10$~G (corresponding to a linear Zeeman splitting $\leq7$~MHz). The frequency $f_\mathrm{{0}}$ corresponds to the splitting of the ground-state manifold in the absence of a magnetic field, so serves as a convenient reference point. The lower x-axis of Fig.~\ref{fig3:sweeps}a indicates the timing of the applied frequency sweep, and the upper x-axis displays the corresponding detuning from $f_0$. The clock transition, $\ket{1,0} \leftrightarrow \ket{2,0}$, experiences no first order Zeeman shift so occurs at $\mathrm{\Delta}_0=0$. In contrast, the transitions above or below this are shifted up or down in frequency, respectively.
The dispersive signal recorded during this frequency sweep displays sharp discontinuities where a state transfer from $F=2$ to $F=1$ (or vice versa) occurs, and we can hence follow the sequential transitions through all of the Zeeman sub-states. Figure~\ref{fig3:sweeps}b shows the dispersive probing signal produced for the same frequency sweep in the presence of an external magnetic field of approximately twice the magnitude, $B_z = 9.3$~G. In this case, the Zeeman splitting of the hyperfine states is larger so the resonances occur at different microwave frequencies. The dotted grey line shows the position of the zero-field $F=1 \leftrightarrow F=2$ transition and the dash-dotted red line indicates what the resultant quantum state would be if the sweep was terminated at $\mathrm{\Delta_{0}}=+11$~MHz. In the 4.7~G field there would not yet have been any transition, meaning the atoms would still populate the $\ket{2,2}$ state, while in the presence of the 9.3~G field, two resonances have been covered and the atoms have been transferred to the $\ket{2,1}$ state. Thus, depending on the magnetic field, different final states are reached using the same microwave field sweep. However, Fig.~\ref{fig3:sweeps} makes it clear that the dispersive probing signal might provide a route to reliably and deterministically prepare states using ARP by providing information about the quantum state population in real-time. It should be noted that as the state population progresses through the Zeeman levels the contrast in the dispersive signal decreases. This is likely due to a combination of atoms leaking into the horizontal dipole trap beam as they escape the finite $z$-confinement of the vertical beam (causing an offset in the signal \cite{Sawyer2017}) and incomplete population transfer via ARP at some stages. The adiabaticity condition for ARP is given by $\Omega_0^2 \gg d/dt (\Delta\omega(t))$ \cite{Malinovsky2001}, where $\Omega_0$ is the Rabi frequency of the transition and $d/dt (\Delta\omega(t))$ is the ARP sweep rate. The Rabi frequencies of the transitions were not directly measured, so imperfect transfer may be a result of non-adiabaticity. However, the lack of contrast in the dispersive signal was not a limiting factor for the work presented in this paper, and in fact highlights the robustness of our feedback algorithm in the presence of noisy input signal. We note that higher-quality data can be acquired with this system \cite{Sawyer2013}.


\begin{figure*}
	\centering
	\includegraphics[width=\textwidth]{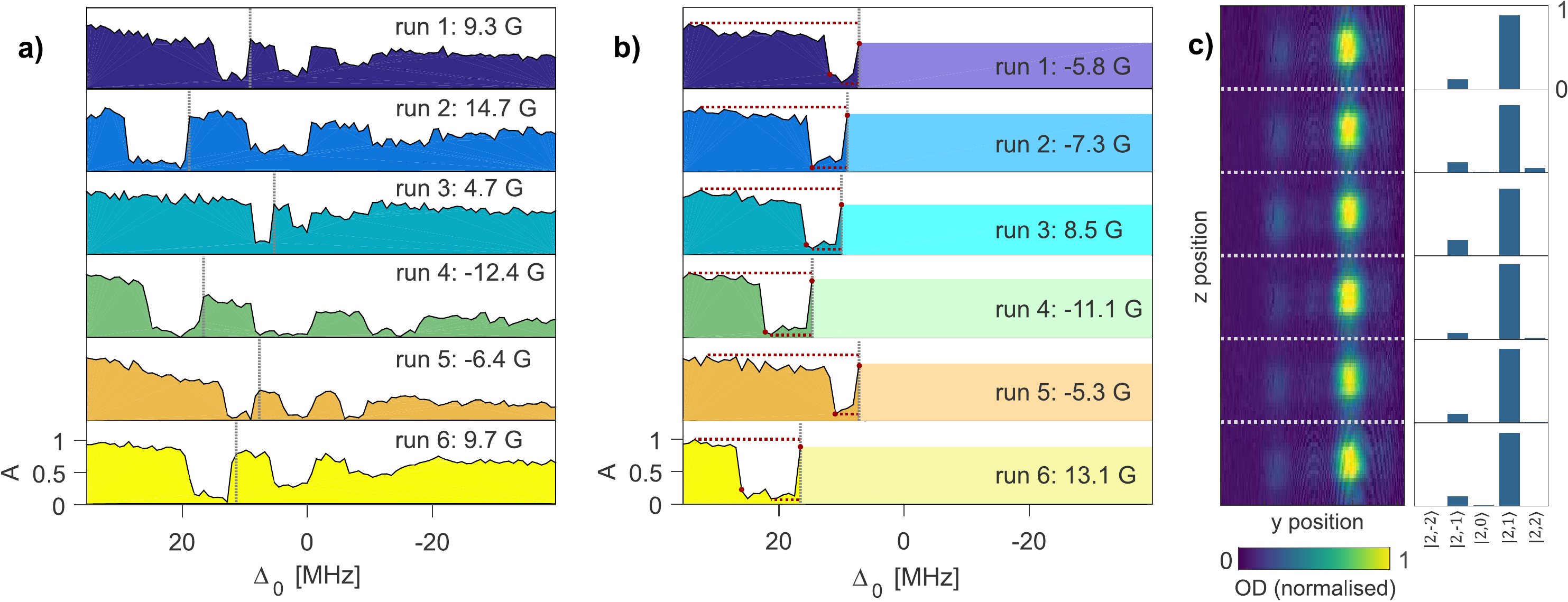}
	\caption{a) Dispersive signal acquired during six separate realisations of a 75~MHz wide driving field sweep centred on $f_0$, applied to an atomic sample in the presence of six different external magnetic fields. b) Data acquired under the same conditions, but with the FPGA feedback loop engaged and set to terminate the drive field after two transitions. The dashed red lines indicate the initial dispersive probing signal magnitude, while the dotted red lines indicate $50\%$ of this value. c) Left: Absorption images corresponding to the six data sets in b, showing the final Zeeman state distributions. Note that the direction of acceleration due to gravity is to the right in the figure. Right: The relative number of atoms in each of the possible Zeeman sub-states, extracted from absorption data.} \label{fig4:resultsDP}
\end{figure*}

To follow the dispersive probing signal in real-time we use an analog-to-digital converter (ADC; Texas Instruments ADS62P49) on board a field-programmable gate-array (FPGA; Xilinx `Virtex 6') to read-in the $I$ and $Q$ output signals from our mixer. The FPGA adds the two signals in quadrature to give the dispersive signal, $A=\sqrt{I^2+Q^2}$ and tracks the minimum and maximum values of $A$ between transitions (initially taking the first measurement and zero to be the maximum and minimum). The algorithm then detects $\geq50\%$ variations in the real-time signal, which indicate the falling and rising edges in $A$ that result from transitions between the $F=1$ and $F=2$ levels. The FPGA program counts these edges, each of which corresponds to a specific transition, and we can produce a particular quantum state by setting the FPGA stop condition to be the number of transitions required to reach that state. When the stop condition is met the FPGA board sends a trigger to the rf switch, terminating the state transfer sequence. A feedback signal is also sent to the switch controlling the AOM that produces the dispersive probing pulse train, such that the probe light is switched off and does not interact any further with the atomic sample.

\section{Results}

As a proof-of-concept demonstration, we use our setup to reliably prepare atoms of a particular quantum state in the presence of a randomised magnetic field. We choose $\ket{2,1}$ as our target state here, though we have tested the system with other target states. We randomly select a bias magnetic field magnitude in the range $\pm (\text{4.5--14.5})$~G and apply it to our sample in order to cause a Zeeman shift of unknown magnitude. We then apply microwave frequency radiation and sweep the frequency at a rate of $-0.3$~MHz/ms across a 75~MHz range centred on $f_{\mathrm{0}}$ while dispersively monitoring the population in the $F=2$ quantum state. Figure~\ref{fig4:resultsDP}a and \ref{fig4:resultsDP}b show the dispersive probing signals acquired during six realisations of this routine, \textit{without} the FPGA feedback system engaged and \textit{with} the FPGA system engaged, respectively. The external magnetic field applied in each case is listed to the right of the corresponding data set. In Fig.~\ref{fig4:resultsDP}a it is evident that the frequency sweep required to reach the $\ket{2,1}$ state -- indicated by the dotted lines on each plot -- varies significantly depending on the external magnetic field.

When the feedback loop is engaged, the FPGA program consistently detects when exactly two transitions have occurred (based on changes in dispersive signal, as discussed in the previous section) and switches off the driving field, leaving our atomic sample prepared in the target state, $\ket{2,1}$. The dotted red lines on the plots of Fig.~\ref{fig4:resultsDP}b indicate the maximum and minimum reference signal levels inferred and used by the FPGA program to determine when the second transition has occurred, and the red circles show the data points that signal each of the first and second transitions to the FPGA. 
There is no dispersive data available after the target state has been reached, but the lighter coloured shaded boxes indicate that the sample remains prepared in the $\ket{2,1}$ state.

The dispersive probe is sensitive to all states in the $F=2$ level, so we need to verify that the $\ket{2,1}$ Zeeman sub-state has indeed been prepared. To quantify the state distribution of the atomic ensemble, we release it into time-of-flight free-fall for 20~ms and make use of the standard Stern-Gerlach separation technique to spatially separate Zeeman sub-levels based on their $m_F$ value before acquiring an absorption image (see App.~\ref{Appendix_SG} for details). Figure.~\ref{fig4:resultsDP}c shows absorption images corresponding to each of the data sets of Fig.~\ref{fig4:resultsDP}b, alongside a bar graph of the relative populations in each of the five possible $y$-positions. This data shows both the $F=1$ and $F=2$ states imaged together, so there is some potential ambiguity in the state composition of the central three bins (the expected $y$-positions of the three $F=1$ sub-states are shown in Fig.~\ref{fig5_SG}b, and compared to a typical absorption image, Fig.~\ref{fig5_SG}c). To check this, we also imaged the two separately to verify that the ensemble with the greatest optical depth consists of only $\ket{2,1}$. The clouds of impurities consist of small amounts of both $\ket{1,1}$ and $\ket{2,-1}$ (top position in Fig.~\ref{fig5_SG}b), $\ket{2,0}$ (second-from-top position in Fig.~\ref{fig5_SG}b), and $\ket{2,2}$ (lower position in Fig.~\ref{fig5_SG}b). 

By analysing the absorption data of Fig.~\ref{fig4:resultsDP}c, we find that $\sim 83\%$ of the initial $\ket{2,2}$ state population is prepared in the $\ket{2,1}$ state when the feedback system is employed.

\section{Discussion}
In the above we have provided a rudimentary experimental realization of our proposal to efficiently transfer atoms to a target quantum state in the presence of an unknown external magnetic field, using a feedback mechanism. Arguably, the field variations that we mitigate in our demonstration are beyond what is typically encountered in a controlled laboratory environment. Hence, relevant applications of the proposed tool are perhaps more likely to be found in a different setting such as field portable devices, on-board experiments, or in a space bourne apparatus harnessing cold atoms \cite{Tino2002,Theron2014}. Here, a limited time window for conducting experiments, violent and uncontrollable environmental conditions, and the requirements of small form factors could motivate the inclusion of a dispersive optical probe as a diagnostic system. In particular, the information stream generated by the the dispersive probe would be highly attractive in the optimization of such experiments subsequent to deployment via machine learning approaches \cite{Wigley2016a}.

Our dispersive probing scheme has a small, but observable, effect on the population dynamics of the atomic system. In \cite{Deb2013} we discuss the effect of photo-spontaneous scattering out of the probe mode and differential light shift from a similar dispersive probing scheme, while other effects include heating, leading to atoms leaking out into the horizontal waveguide trapping beam (the recoil energy of a $^{87}$Rb atom scattering a photon in the D2 line is $181~\mathrm{nK} \times k_{B}$), and decoherence, where some atoms return to the ground state. To quantitatively estimate the effect of optical pumping from the dispersive probe triplet we solved the optical Bloch equations for the multi-level system of Fig.~\ref{fig1:setstage}b, where two sets of dispersive probe pulses were applied to a sample of atoms initially prepared in $\ket{2,2}$\footnote{We made use of the open-source `{AtomicDensityMatrix}' package for Mathematica: see rochesterscientific.com/ADM.}. At the end of each set of pulses, an ARP state transfer was used to swap the populations of atoms $\ket{2,2} \leftrightarrow \ket{1,1}$ (following pulse set one) and $\ket{1,1} \leftrightarrow \ket{2,1}$ (following pulse set two). The maximum number of probe pulses applied before the first microwave transition was 20 in the experiments presented here, so we used 20 pulses in each set to account for the `worst case' scenario. The estimated final population distribution was $89\%$ in $\ket{2,1}$, $5.5\%$ in $\ket{1,1}$, $4.5\%$ in $\ket{2,2}$, $0.5\%$ in $\ket{2,0}$, and $0.5\%$ in $\ket{1,0}$. This compares favourably with the $83\%$ transfer to $\ket{2,1}$ that we observe in the data of Fig.~\ref{fig4:resultsDP}c.

To obtain purer end-states, resonant light pulses could be incorporated into the sequence to remove any atoms left behind in $F=2$ each time the sample is transferred to the $F=1$ level, and vice-versa, and the ARP sweep rate could be further optimised.
Additionally, dispersively probing with higher resolution would allow us to more precisely detect the transition points, which would lead to the method being viable when the external magnetic field is smaller than $\pm4.5$~G. However, increasing the dispersive probe pulse frequency (or power) results in more interaction with the atomic sample, which may cause transfer to unwanted impurity states and atom loss. The effect of the dispersive probe field could be completely eliminated by augmenting the scheme with additional ensembles \cite{Chisholm2018}. In this way information can be gathered by probing small auxiliary clouds without affecting the main cloud at all. 

It should be noted that in the presence of a temporally constant background magnetic field, as has been used to demonstrate the method in this paper, we only, at least in principle, have to probe and detect the first transition to determine the Zeeman splitting, and thus where to terminate the sweep for any of the following states on the ladder. If the magnetic field were time-dependent, however, we would need to continue to probe the state population throughout the entire sweep as we have done here.

In the future, we plan to utilise the feedback system to prepare specific superposition states through adaptive measurement, where dispersive probing during an ARP sweep is used to first learn the resonant frequency, then resonant pulses used to prepare the target state.

\section{Conclusion}

We have built a field-programmable gate-array (FPGA) based feedback loop that samples and analyses an off-resonant optical probe signal and sends an output signal that terminates the sweeping microwave radiation field when the target $\ket{F,m_F}$ state has been achieved. The system has been applied to reliably transfer atoms to a specific quantum state ($\ket{2,1}$) in the presence of a random magnetic field in the range $\pm (4.5-14.5)$~G.

\section*{Author contribution statement}

B.J.S and M.C implemented the scheme and performed the experiments with support from R.T and A.B.D. B.J.S. analysed the data. B.J.S. and A.B.D. performed the simulations. B.J.S. and N.K. prepared the manuscript with input and comments from all authors. N.K. supervised the project.

\begin{acknowledgement}
We thank Jevon Longdell and Daniel Schumayer for critically reading our manuscript.
We also acknowledge Jeremy Lee-Hand and Daniel Schumayer for their help on setting up the FPGA board and for discussions on
feedback loops.
\end{acknowledgement}
	
\appendix

\section{Stern-Gerlach separation technique} \label{Appendix_SG}

\begin{figure}[!htbp]
	\centering
	\resizebox{1\columnwidth}{!}{\includegraphics[trim=0cm 0cm 0cm 0cm]{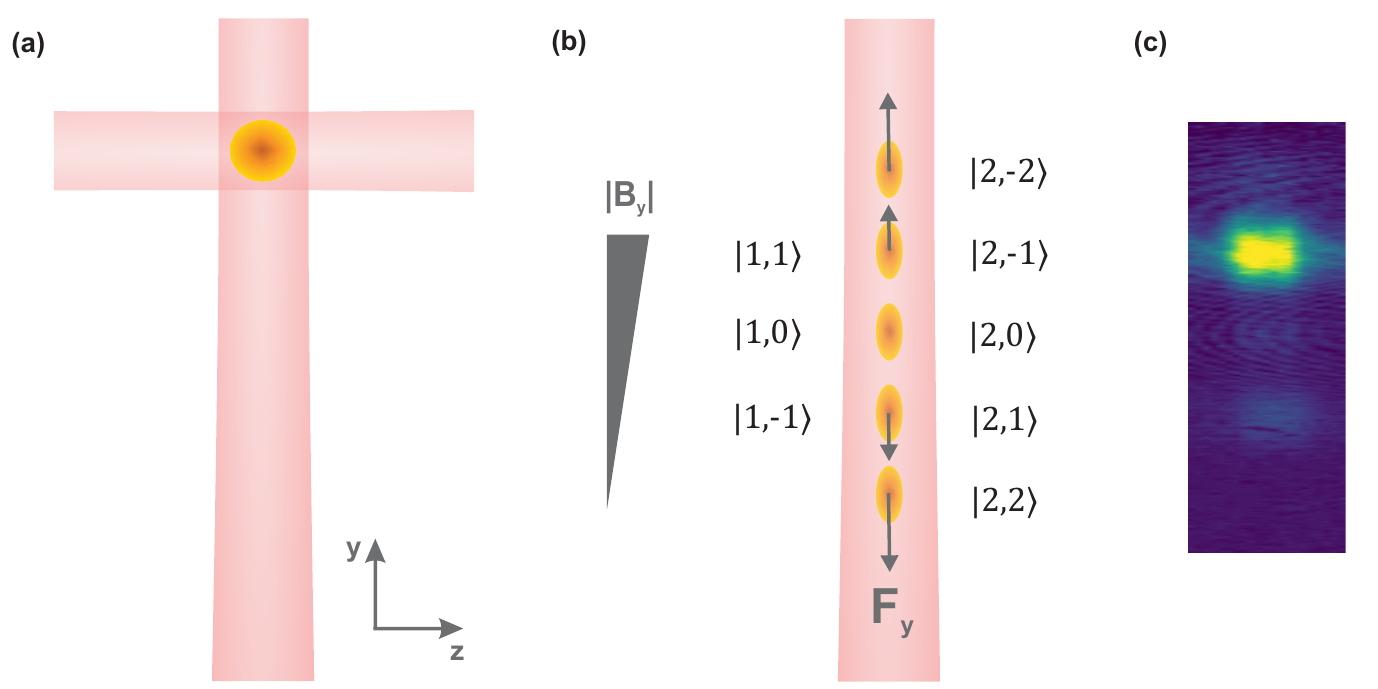}}
	\caption{a) Schematic of the atomic sample trapped in the cross-beam dipole trap. b) Atomic sample distribution at the end of a 20~ms time-of-flight showing all possible states, where states have been separated out based on their $m_F$ number in the presence of a magnetic field gradient along $y$. c) The third panel of Fig.~\ref{fig4:resultsDP}c, showing an absorption image of the final Zeeman state distribution following ARP and dispersive probing. The panel is rotated relative to Fig.~\ref{fig4:resultsDP}c so the orientation matches that of Fig.~\ref{fig5_SG}b.} \label{fig5_SG}
\end{figure}

We release the atomic ensemble from the dipole trap (see Fig.~\ref{fig5_SG}a) into time-of-flight free-fall for 20~ms. An inhomogeneous magnetic field 
is applied to the mixed-state ensemble during free-fall for the first 10~ms. Each atom is accelerated by an additional magnetic force
\begin{equation}
\mathbf{F_y }= -g_f m_F \mu_B \frac{\partial \mathbf{B_y}}{\partial y},
\end{equation}
where $\mu_B$ is the Bohr magneton and $g_F$ is the Land\'e $g$-factor (with $g_F\simeq1/2$ for $F=2$, and $g_F\simeq-1/2$ for $F=1$).
This results in separate atomic clouds at five different possible positions along the $y$-axis, depending on the hyperfine state distribution, as illustrated schematically in Fig.~\ref{fig5_SG}b. Note that because $g_f$ has the same magnitude for both hyperfine levels, there are three instances where both an $F=1$ and $F=2$ population are displaced to the same position. In order to maintain a high signal-to-noise ratio we leave the vertical dipole trapping beam on while the states are separating out, so we do not get too much spreading in the $z$-direction.

At the end of the 20~ms time-of-flight we switch off the vertical trapping beam and use light resonant with the $F=2 \rightarrow F'=3$ transition to acquire an absorption image of the sample. If we want to observe atoms in $F=1$ states visible also, we first transfer them to the $F=2$ level using a pulse of repump light, resonant with the $F = 1 \rightarrow F' = 1-2$ cross-over resonance, before we take the absorption image.


\begin{thebibliography}{10}
	\providecommand{\url}[1]{{#1}}
	\providecommand{\urlprefix}{URL }
	\expandafter\ifx\csname urlstyle\endcsname\relax
	\providecommand{\doi}[1]{DOI \discretionary{}{}{}#1}\else
	\providecommand{\doi}{DOI \discretionary{}{}{}\begingroup
		\urlstyle{rm}\Url}\fi
	
	\bibitem{Abragam1961}
	P.~Abragam, A.~Abragam,
	\newblock \emph{The Principles of Nuclear Magnetism}
	\newblock  (Clarendon Press, 1961)
	
	\bibitem{Allen1987}
	L.~Allen, J.H. Eberly,
	\newblock \emph{{Optical Resonance and Two-Level Atoms}} (Dover Publications,
	1987)
	
	\bibitem{Camparo1984}
	J.C. Camparo, R.P. Frueholz, J. Phys. B (\textbf{17}), 4169
	\newblock  (1984)
	
	\bibitem{Kotru2014}
	K.~Kotru, J.M. Brown, D.L. Butts, J.M. Kinast, R.E. Stoner, Phys. Rev. A
	(\textbf{90}), 053611 (2014)
	
	\bibitem{Malinovsky2001}
	V.~Malinovsky, J.~Krause, Eur. Phys. J. D (\textbf{14}), 147 (2001)
	
	\bibitem{Bjorklund1980}
	G.C. Bjorklund, Opt. Lett. (\textbf{5}), 15
	\newblock  (1980)
	
	\bibitem{Lye1999}
	J.E. Lye, B.D. Cuthbertson, H.A. Bachor, J.D. Close, J. Opt B (\textbf{1}), 402
	(1999)
	
	\bibitem{Sawyer2012}
	B.J. Sawyer, A.B. Deb, T.~McKellar, N.~Kj\ae{}rgaard, Phys. Rev. A
	(\textbf{86}), 065401
	\newblock  (2012)
	
	\bibitem{Kohnen2011}
	M.~Kohnen, P.G. Petrov, R.A. Nyman, E.A. Hinds, New J. Phys. (\textbf{13}),
	085006
	\newblock  (2011)
	
	\bibitem{Bason2018}
	M.G. Bason, R.~Heck, M.~Napolitano, O.~El{\'{\i}}asson, R.~M{\"{u}}ller,
	A.~Thorsen, W.Z. Zhang, J.J. Arlt, J.F. Sherson, J. Phys. B: At., Mol. Opt.
	Phys. (\textbf{51}), 175301 (2018)
	
	\bibitem{Smith2004}
	G.A. Smith, S.~Chaudhury, A.~Silberfarb, I.H. Deutsch, P.S. Jessen, Phys. Rev.
	Lett. (\textbf{93}), 163602 (2004)
	
	\bibitem{Isayama1999}
	T.~Isayama, Y.~Takahashi, N.~Tanaka, K.~Toyoda, K.~Ishikawa, T.~Yabuzaki, Phys.
	Rev. A (\textbf{59}), 4836
	\newblock  (1999)
	
	\bibitem{Deb2013}
	A.~Deb, B.J. Sawyer, N.~Kj\ae{}rgaard, Phys. Rev. A (\textbf{88}), 063607
	\newblock  (2013)
	
	\bibitem{Gajdacz2016}
	M.~Gajdacz, A.J. Hilliard, M.A. Kristensen, P.L. Pedersen, C.~Klempt, J.J.
	Arlt, J.F. Sherson, Phys. Rev. Lett. (\textbf{117}), 073604
	\newblock  (2016)
	
	\bibitem{Vanderbruggen2013}
	T.~Vanderbruggen, R.~Kohlhaas, A.~Bertoldi, S.~Bernon, A.~Aspect, A.~Landragin,
	P.~Bouyer, Phys. Rev. Lett. (\textbf{110}), 210503
	\newblock  (2013)
	
	\bibitem{Vanderbruggen2014}
	T.~Vanderbruggen, R.~Kohlhaas, A.~Bertoldi, E.~Cantin, A.~Landragin, P.~Bouyer,
	Phys. Rev. A (\textbf{89}), 063619 (2014)
	
	\bibitem{Gillett2010}
	G.G. Gillett, R.B. Dalton, B.P. Lanyon, M.P. Almeida, M.~Barbieri, G.J. Pryde,
	J.L. O'Brien, K.J. Resch, S.D. Bartlett, A.G. White, Phys. Rev. Lett.
	(\textbf{104}), 080503 (2010)
	
	\bibitem{Dong2010}
	D.~Dong, I.~Petersen, {IET} Control Theory {\&} Applications (\textbf{4}), 2651
	(2010)
	
	\bibitem{Serafini2012}
	A.~Serafini, {ISRN} Optics (\textbf{2012}), 1 (2012)
	
	\bibitem{Steck2004}
	D.A. Steck, K.~Jacobs, H.~Mabuchi, T.~Bhattacharya, S.~Habib, Phys. Rev. Lett.
	(\textbf{92}), 223004 (2004)
	
	\bibitem{Mabuchi2008}
	H.~Mabuchi, Phys. Rev. A (\textbf{78}), 032323 (2008)
	
	\bibitem{Wilson2007}
	S.D. Wilson, A.R.R. Carvalho, J.J. Hope, M.R. James, Phys. Rev. A
	(\textbf{76}), 013610 (2007)
	
	\bibitem{Szigeti2010}
	S.S. Szigeti, M.R. Hush, A.R.R. Carvalho, J.J. Hope, Phys. Rev. A
	(\textbf{82}), 043632
	\newblock  (2010)
	
	\bibitem{Wang2016}
	S.~Wang, T.~Byrnes, Phys. Rev. A (\textbf{94}), 033620 (2016)
	
	\bibitem{Hentschel2010}
	A.~Hentschel, B.C. Sanders, Phys. Rev. Lett. (\textbf{104}), 063603 (2010)
	
	\bibitem{Roberts2014}
	K.O. Roberts, T.~McKellar, J.~Fekete, A.~Rakonjac, A.B. Deb, N.~Kj\ae{}rgaard,
	Opt. Lett. (\textbf{39}), 2012
	\newblock  (2014)
	
	\bibitem{Sawyer2017}
	B.J. Sawyer, M.S.J. Horvath, E.~Tiesinga, A.B. Deb, N.~Kj\ae{}rgaard, Phys.
	Rev. A (\textbf{96}), 022705
	\newblock  (2017)
	
	\bibitem{Bize1999}
	S.~Bize, Y.~Sortais, M.S. Santos, C.~Mandache, A.~Clairon, C.~Salomon,
	Europhysics Letters ({EPL}) (\textbf{45}), 558 (1999)
	
	\bibitem{Sawyer2013}
	B.J. Sawyer, Dispersive probing of quantum state preparation in ultracold
	$^{87}${Rb}.
	\newblock Master's thesis, University of Otago
	\newblock  (2013)
	
	\bibitem{Tino2002}
	G.~Tino, Nucl. Phys. B Proc. Suppl. (\textbf{113}), 289 (2002)
	
	\bibitem{Theron2014}
	F.~Theron, O.~Carraz, G.~Renon, N.~Zahzam, Y.~Bidel, M.~Cadoret, A.~Bresson,
	Appl. Phys. B (\textbf{118}), 1 (2014)
	
	\bibitem{Wigley2016a}
	P.B. Wigley, P.J. Everitt, A.~van~den Hengel, J.W. Bastian, M.A.
	Sooriyabandara, G.D. McDonald, K.S. Hardman, C.D. Quinlivan, P.~Manju, C.C.N.
	Kuhn, I.R. Petersen, A.N. Luiten, J.J. Hope, N.P. Robins, M.R. Hush,
	Scientific Reports (\textbf{6}) (2016)
	
	\bibitem{Chisholm2018}
	C.S. Chisholm, R.~Thomas, A.B. Deb, N.~Kj{\ae}rgaard, Rev. Sci. Instrum.
	(\textbf{89}), 103105 (2018)
	
\end{thebibliography}

\end{document}